# Revisiting the Impact of Single-Vacancy Defects on Electronic Properties of Graphene

Mohammadamir Bazrafshan*, Thomas. D. Kühne

*Center for Advanced Systems Understanding (CASUS), Am Untermarkt 20, 02826 Görlitz, Germany*

*Corresponding author's email: Mohammadamirbazrafshan@gmail.com*

**Abstract:** While defects are generally considered to be unavoidable in experiments, engineering them is also a way of manipulating the physical properties of materials. In this study, the role of periodically arranged single vacancy defects in graphene is studied using the tight-binding method. Our numerical results show that single vacancy (SV) defects can exhibit predictable electronic behavior when they reside on the same sublattices (SS), following the armchair graphene nanoribbons (AGNRs) electronic band structure depending on the spacing between SVs. AGNRs are known to their tunable electronic band gap. However, when they are located on different sublattices (DS), the interaction between the defect-induced states becomes strong and can introduce anisotropy into the electronic band structure, demonstrating that the relative position of the SVs can also act as an additional degree of freedom for tuning the electronic properties. Interestingly, the behavior is independent of the density of SVs; for a system fully defected with SVs, the electronic properties depend heavily on the sublattices involved. The results provide a novel insight into sublattice-based defect engineering.

## I. Introduction

In 2004, a single layer of graphite was obtained through mechanical exfoliation, which revolutionized the field of material engineering and opened up new possibilities in the study of low-dimensional materials [1]. In 1947, the main electronic properties of graphene were modeled using a tight-binding approach, which predicted its zero-band gap [2]. The model implies that graphene's electrical conductivity is weakly temperature-dependent due to the small number of thermally excited carriers. The linear bands at Dirac points in graphene's electronic band structure originated from $p_z$ orbital provide a very rich physics, from half quantum integer Hall effect [3], to exceptionally high electronic conductivity [4] despite its vanishing density of states close to the Fermi energy. The stability of graphene in conventional environments and its fascinating physical properties have brought a huge research interest [5,6].

From a crystallographic perspective, graphene is a single layer of carbon atoms arranged in a honeycomb lattice consisting of two equivalent sublattices in a triangular geometry. Any imbalance or symmetry breaking between the sublattices can induce significant lattice distortion and profoundly alter the electronic and physical properties of the system. Producing a perfect crystal is challenging [7]. However, defect engineering is also a way to manipulate physical properties. In this respect, the effect of various defects has been studied extensively in the literature, including single vacancies [8–11], divacancies [12,13], Stone Wales [8,14,15] and impurity atoms [16–18]. One of the structural defects studied and observed experimentally [19] is the single vacancy, for which a general picture is presented and discussed in the literature. The picture is based on the bipartite nature of graphene, which states that SVs on the same sublattice induce ferromagnetic ordering and a band gap. SVs on both sublattices, however, can lead to antiferromagnetic behavior and the absence of a band gap [20]. However, the reports in the literature do not seem to be consistent from the electronic properties point of view [20–23].

In order to gain a better understanding of the effect of SVs on the sublattices involved, we systematically studied SVs that were arranged periodically and located on the same or different sublattices. Our numerical simulations based on the tight-binding (TB) approximation show that, depending on their spacing along the zigzag direction, the electronic properties of SVs on the same sublattice exhibit a repeating pattern similar to that of AGNRs. The electronic properties of a N-AGNR are width dependent and can be classified into three families depending on their width N, where N is the number of carbon atoms across the ribbon. These families exhibit a periodic modulation of the band gap, repeating every three atoms, leading to distinct metallic (gapless) or semiconducting behavior depending on whether N=3p, 3p+1, or 3p+2, where p is an integer [24,25]. However, zigzag graphene nanoribbons (ZGNRs) are always metallic [26]. These features can be well captured using the TB approximation.

Moreover, for SVs located on the same sublattice there always flat bands corresponding the number of SVs that is an outcome of linear algebra in numerical calculations [21,27], but the band gap (neglecting flat bands) can be modulated by controlling the distance between. On the other hand, when SVs are distributed on different sublattices, the electronic band structure shows a complex behavior. Nevertheless, the two bands close to the Fermi energy are not as flat as the case in SS, showing dispersion while they are not mixed with the band continuum. The Fermi velocity is also analyzed and it is shown in some cases the defected crystal can present anisotropy for electronic properties.

The manuscript is organized as follows: The next section presents the geometry and theoretical approach. Section III discusses the numerical results. Section IV concludes our study.

## II. Model and Method

The atomic structure of a graphene sheet is shown in Figure 1 (a), with the two sublattices in different colors. The size of the unit cells is changed to be able to study other possible structures. To build a rectangular unit cell, n and m must be defined as even numbers. We define the unit cell vectors **a** and **b** in the X and Y directions of the Cartesian coordinate system. In panel (b) a (n=16, m=6) unit cell is shown. Also, in the pure case, the removed atom(s) can be addressed using m and n, where n controls the spacing along the zigzag direction of the graphene lattice and m controls the spacing along the armchair direction. The naming convention is illustrated in Figure 1 (b). If all the removed atoms belong to the same sublattice, the model is labelled SS, otherwise, it is labelled DS. The spacing between SVs is included in the model's name, depending on how many SVs exist in the unit cell. If there is only one SV, a single value appears; if there are multiple SVs, multiple spacing values are added accordingly. For brevity, we only mention the spacing along the X axis. The unit cell sizes for each system are not mentioned.

The electronic properties of pure and vacancy defected graphene systems are studied by employing $p_z$ orbital real-space tight-binding model with the Hamiltonian that reads:

$$H = -t \sum_{\langle ij \rangle}(|i\rangle\langle j| + |j\rangle\langle i|) \quad (1)$$

with i and j indices indicate the label of each atom, and t is the hopping integral. The onsite energy is considered zero, and not included in the formula. The hopping integral is considered $-2.7$ eV. The density of states from the band structure is calculated using Gaussian smearing, $\text{DOS}(E) = \sum_l \sum_{k \in BZ} \frac{1}{\sigma\sqrt{\pi}} e^{-\frac{(E-E_l(k))^2}{\sigma^2}}$, with $l$ as band index , $k$ as the wave vector, and $\sigma$ is the Gaussian broadening with a width if $0.01$ eV. The resulting DOS is in arbitrary units. The Fermi velocity is calculated for the valence band using $v_F = \frac{1}{\hbar}|\frac{dE}{dk}|$, where when needed, the k-point is indicated by a superscript for a specific k-path.

Interested readers can use the Python code to form pure or defective graphene superlattices according to the notation in the manuscript, and calculate the TB band structure. All figures can be generated using the numbers referenced as {[Number]} by the code.

First-principles DFT calculations were performed using the SIESTA 5.4.0 package [28], with a double-ζ basis set and the GGA using the PBE functional [29] which offer a good balance between accuracy and computational cost [30]. An energy shift of 0.03 eV and a real-space grid equivalent to a plane-wave cutoff of 320 Ry were used. The K-point mesh converges with a sampling of 18×18×1 using the Monkhorst-Pack scheme [31].

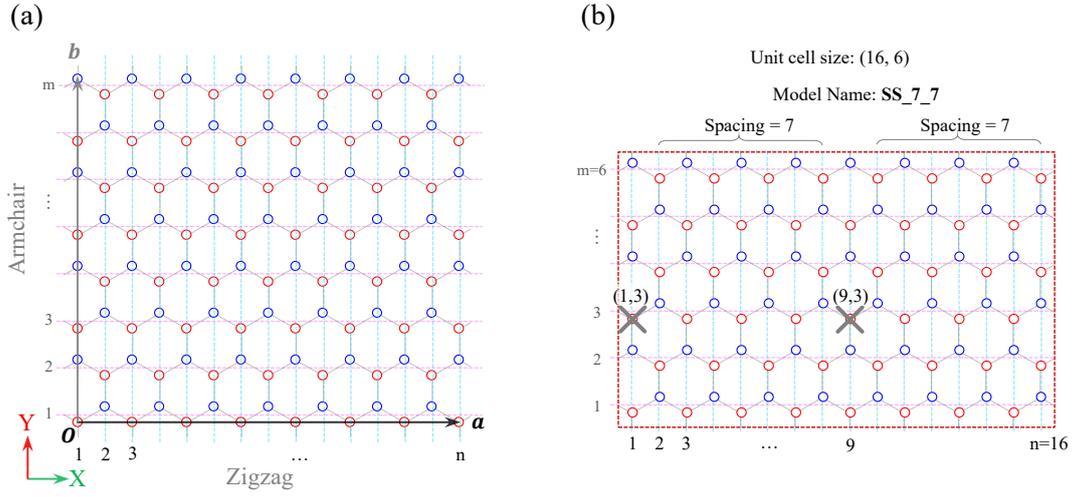

Figure 1. (a) Atomic model of graphene, showing the two sublattices in different colors. (b) An example of a (16,6) unit cell with two vacancies located on the same sublattice.

## III. Results and Discussion

The electronic band structure of graphene obtained from a DFT calculation with a rectangular unit cell is shown in Figure 2. The linear bands meet at the Dirac point, which is indicated along the X direction, corresponding to one of the zigzag directions of the sheet. It is worth mention that the Dirac point can be folded into other points along $\Gamma - X$ of the k-path by increasing unit cell size along X direction of the lattice. In the following two subsections, we study the effect of the occurrence of SVs on the same or different sublattices.

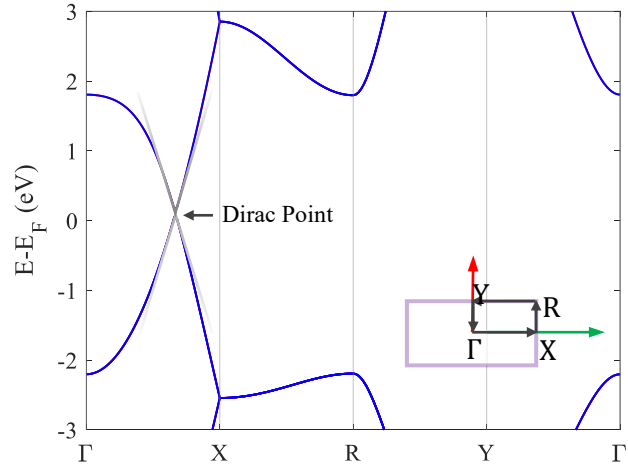

Figure 2. The electronic band structure of perfect graphene with a rectangular unit cell. The first Brillouin zone and the k-path are shown in the inset. Two lines are plotted at the Dirac point to better indicate the linear slope of the bands.

### SVs on the same sublattice

In this subsection we examine the impact of SVs when they are located on the same sublattice. Imbalances between sublattices can drastically alter electronic properties, including the formation of unpaired states that appear as non-dispersive bands in the electronic band structure of a periodic system. However, we first examine the spacing between SVs when there is just one SV in the unit cell. The electronic band structure of SS_3, SS_5 and SS_7 are presented respectively in Figure 3 (a), (b) and (c) {[3]}. Sizing up unit cells along the Y direction does not induce a drastic change in the general behavior of the system, for example SS_5 keeps its touching bands at the folded Dirac point {[31]}.

As it is mentioned earlier, the electronic properties of AGNRs exhibit a well-known repeating pattern, i.e., their electronic band gap follows 3p+2<3p<3p+1 with p as integer. We refer 3p+2 group as gapless hereafter. Interestingly, Figure 3 also suggests that this behavior emerges here, note the spacing between the SVs. In other words, an AGNR with a width of three atoms in the width (3AGNR) is a semiconductor, while 5AGNR is gapless, specifically, SS_5 show a gapless electronic behavior while SVs are placed on the same sublattices.

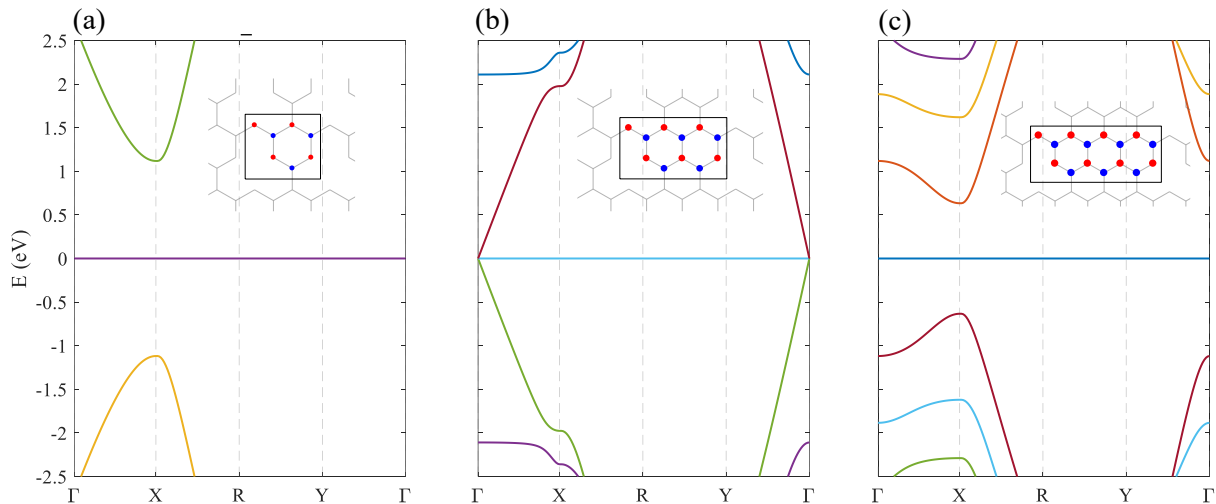

Figure 3. The electronic band structure of (a) SS_3, (b) SS_5, and (c) SS_7 vacancy defected graphene sheet. The unit cells are shown in the insets.

To confirm such a behavior, we consider other models with a greater number of SVs. In Figure 4 (a) the SVs are arranged so that the gapless AGNRs can be considered with different widths. In panels (b) and (c), two and three SVs are placed with spacings that follows the AGNRs with a band gap {[4]}. The results indicate when SVs are placed on the same sublattice, regardless of their density, the electronic properties of the systems follow the one of AGNRs, e.g., SS_2_2 which has the maximum SV concentration exhibit linear bands in its electronic band structure (not shown here, {[41]}). For models containing both gapless and gapped AGNRs, the electronic band structure does not exhibit the gapless feature {[42]}, suggesting extra care is needed when using molecule-based simulations due to the effect of the edges.

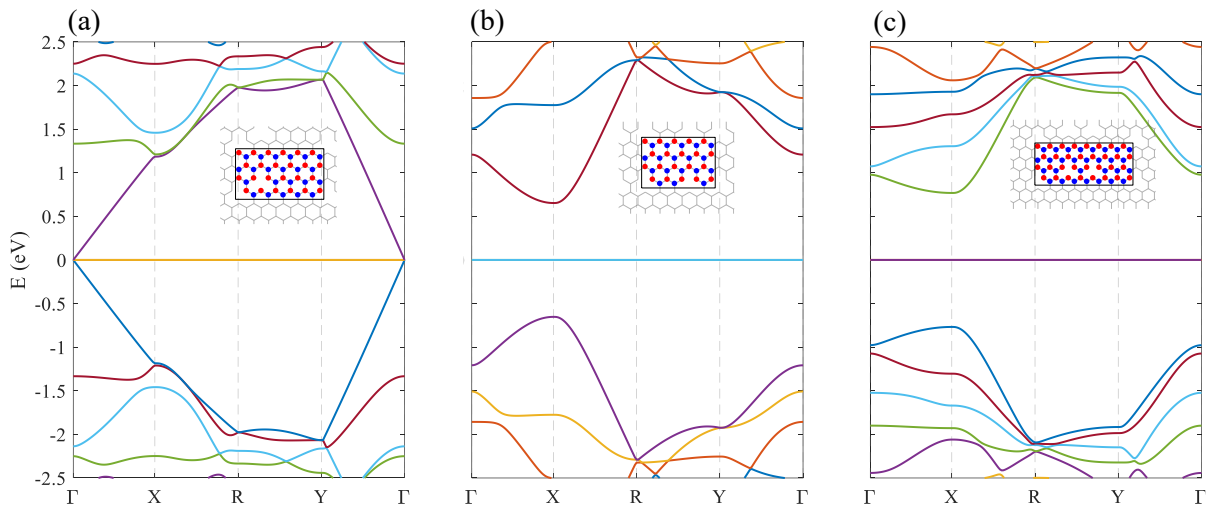

Figure 4. The electronic band structure of (a) SS_2_8, (b) SS_5_3, and (c) SS_4_3_6.

The Fermi velocity is calculated ~$8.7 \times 10^5$ m/s, confirming that the linear bands are those of graphene which has a rich physics. In the next section we study the effect of SVs on different sublattices.

## SVs on Different sublattices

We have studied how behavior of the band structure can be predicted when the SVs are located on the same sublattice. In this subsection we first examine SVs with different spacings along the X direction. In Figure 5, two systems are shown, namely DS_5_5 (a), and DS_6_4 (b), {[5]}. The electronic band structure of both models shows two impurity (defect-induced) bands in which are isolated from the band continuum. However, their behavior along the wavevector differs significantly. Nevertheless, the dispersion of isolated bands can indicate how strongly the vacancy-induced states interact with each other. Comparing these results with those of the SS suggests that these states are quite different, with flat bands becoming dispersive but not extending into the band continuum. While SVs on the SS clearly show AGNR character emergence, those on DSs do not. The two bands close to the Fermi energy in a ZGNR become non-dispersive over a range of k-vectors because of their highly localized nature at the edges.

The DOS plot for DS_6_4 shows a vanishing trend close to the Fermi energy, similar to that of graphene. In other words, analyzing only the DOS can produce results similar to those of graphene, while the band structure can differ. The peak in the DOS plot is mainly due to the small dispersion of the band along the $\Gamma - R$ wavevector path. The DOS of DS_6_4 exhibits higher values, originating from impurity bands whose shapes vary smoothly along the wave vector. It should be noted that the isolated valence and conduction bands are clearly separated from the band continuums by a band gap. However, it is worth noting that when the number of SVs is odd, there are practically two SVs on the SS. This means that the emergence of a flat band in simulations is unavoidable and that the lattice is uncompensated.

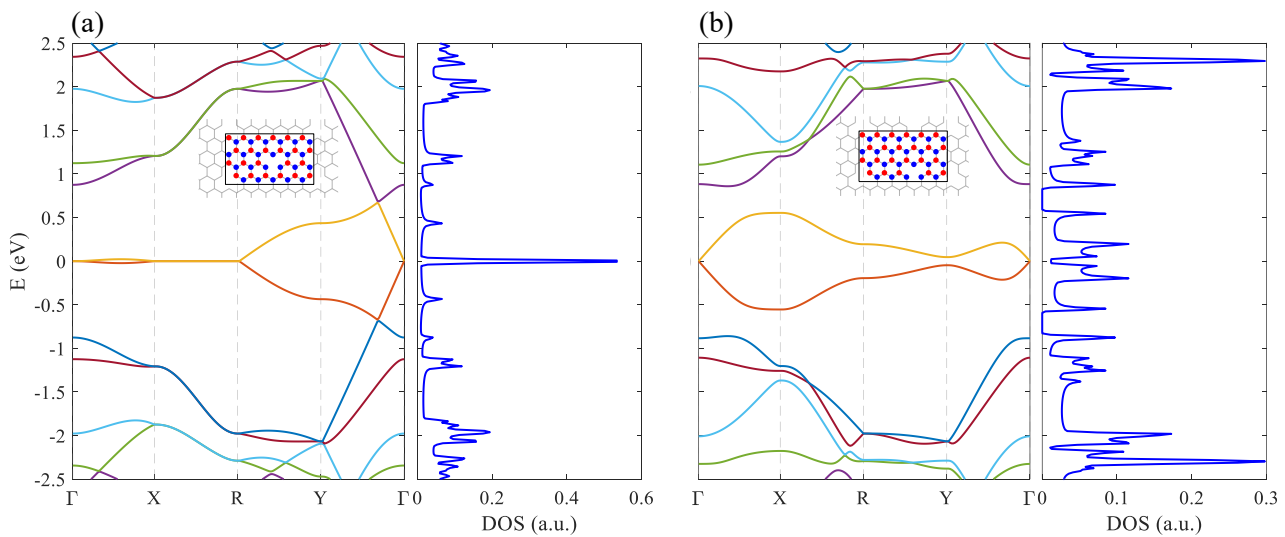

Figure 5. The electronic band structure (left) and DOS plot (right) for (a) DS_5_5, (b) DS_6_4.

As discussed, there is no clear indication of a ZGNR or an AGNR in the band structure. However, in some regions, there is a small band dispersion, while in others, there is a higher band dispersion. This suggests that a variety of effects may be involved. Furthermore, when comparing SVs on the same sublattice, DS systems do not necessarily have a non-dispersive band that corresponds to the number of vacancies. Instead, they can hybridize with each other, resulting in complex behavior. Another interesting feature of the band structure is that the Fermi velocities along the $\Gamma - X$ ($\mathcal{O} \times 10^3$ m/s) and $Y - \Gamma$ ($\mathcal{O} \times 10^4$ m/s) close to the Fermi energy have different values for DS_5_5, suggesting the system shows a strong anisotropy for electronic properties.

In situations where the concentration of SVs is maximum but on a different sublattice (namely DS_2_2, not shown here, {[51]}), the electronic band structure is similar to that of DS_5_5. However, the vacancy defect concentration is equal to that of the case where the SVs are on the same sublattice, or SS_2_2. This shows that the sublattices involved have a profound impact on determining the electronic properties, as well as the spacing between them.

The relative position of SVs by the means of changing m is studied in Figure 6 for cases where just the position of the SVs is changed and the spacing between them, n, kept equal {[6]}. In panel (a), the distance along the Y direction, m, between SVs is the smallest, and (b) has the largest distance. By comparing the results, one can find that the dispersion of two bands close to the Fermi energy along the $\Gamma - X$ path can be changed. Zooming in on the bands reveals that those on the right (toward $\Gamma$) have a smooth shape, while those on the right (toward X along the k-vector) are more linear. This suggests that the bands on the right are similar to those of a ZGNR, whereas those on the left are not. Also, the value of the Fermi velocity shows that $v_F^\Gamma < v_F^X$ are not equal, i.e., $v_F^X$ can has a higher contribution in electronic transport. However, as mentioned earlier, these bands are isolated from the band continuum, meaning they are highly sensitive to hopping terms, which are reflected in bond lengths in real space.

The band dispersion is very complex in cases where SVs are distributed on different sublattices, but changing the distance between SVs along the Y direction of the lattice can also alter the bands close to the Fermi energy, and can even induce a strong anisotropy. It should be noted that the calculated Fermi velocities can only provide qualitative insight. Nevertheless, an AGNR with a width of 5 atoms, for example, has a very small energy gap in DFT simulations, whereas in simple TB it is gapless.

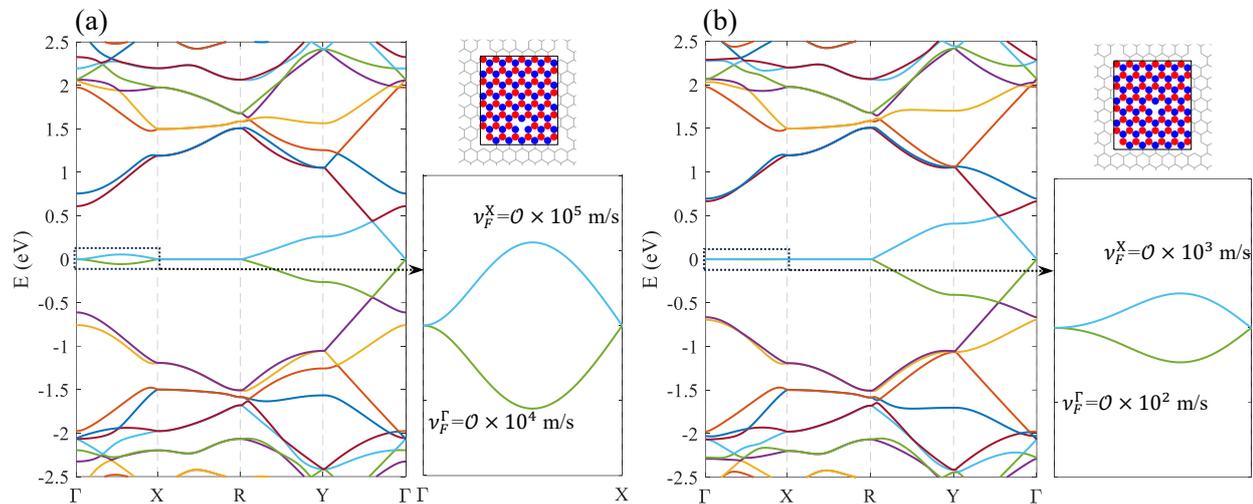

Figure 6. The electronic band structure (left), the unit cell, and a magnified view of the region marked by the black box (right) for (a) DS_5_5 with SVs at the closest Y direction separation. (b) The SVs are placed at the maximal Y distance. The supercell size is (5,5). The Fermi velocity is shown close to the $\Gamma$ and X.

## IV.   Conclusion

We have studied the electronic properties of periodically arranged single vacancies using simple tight-binding method in two cases whether the SVs are located on the same or different sublattices. Analyzing numerical results reveals two distinct behaviors, namely for SVs on the same sublattices the spacing between them determines their electronic properties, following those of an AGNR, while if they are on different sublattices, the behavior is more complex but with two isolated bands close to the Fermi energy. Tuning band gap can be useful in optical and the electronic transport properties. It is shown in some cases it is possible to induce anisotropy in electronic properties by engineering them. Our results show that the electronic band structure is governed by the placement of defects within sublattices, rather than their density, even in systems with the maximum possible SV concentration. The results provide a new insight into single vacancy defects and their effects on electronic properties.

## Code availability

An interactive Colab notebook is provided for reviewer access at:

https://colab.research.google.com/drive/1LuPNap9xzauSEGXA-RaHMuqyNAu8qz6k?usp=sharing